\documentclass[aip,pop,sd,reprint,10pt,amsmath,amssymb,graphicx]{revtex4-1}

\usepackage[capitalize]{cleveref}
\usepackage[T1]{fontenc}
\usepackage{float}
\usepackage{rotating}
\usepackage[normalem]{ulem}
\usepackage{amsmath}
\usepackage{textcomp}
\usepackage{color}
\usepackage{marvosym}
\usepackage{wasysym}
\usepackage{amssymb}
\usepackage{bm}

\newcommand{\beq}{\begin{equation}}
\newcommand{\eeq}{\end{equation}}
\newcommand{\bfig}{\begin{figure}[htbp]}
\newcommand{\efig}{\end{figure}}

\newcommand{\ben}{\begin{eqnarray}}
\newcommand{\een}{\end{eqnarray}}
\usepackage{graphicx}
\usepackage{color}
\usepackage{subfig}
\usepackage{subfig}

\usepackage{epsfig}
\usepackage[titletoc,toc,title]{appendix}

\begin{document}
\title{Coherent  transport structures  in magnetized plasmas\,   {\it I}  : Theory}
\author{G. Di Giannatale}\affiliation{IGI - CNR Corso Stati Uniti 4, Padova, Italy}
\author{M.V. Falessi}\affiliation{Dipartimento di Matematica e Fisica,  Roma Tre University,  Roma, Italy}
\author{D. Grasso}\affiliation{ISC - CNR and Politecnico di Torino\\  Dipartimento Energia C.so Duca degli Abruzzi 24, Torino. Italy }
\author{F. Pegoraro}\affiliation{Dipartimento di Fisica E. Fermi,  Pisa University, \\ largo Pontecorvo 3, Pisa, Italy}
\author{T.J. Schep}\affiliation{Fluid Dynamics Laboratory, Dep. Applied of Physics, Eindhoven University of Technology, p.o.box 513, 5600MB Eindhoven, The Netherlands}

\bigskip

\begin{abstract}
{In a pair of linked  articles (called Article {\it I} and  {\it II}  respectively) we apply the concept of Lagrangian Coherent Structures {(LCSs)} borrowed from the study of Dynamical Systems to magnetic field configurations in order to separate regions where field lines have different kind of behaviour.
\\  In  the present article, article {\it  I}, after recalling the definition and the properties of  the {LCSs}, we show how  this  conceptual framework  can be applied to the study of particle transport in a magnetized plasma.  Futhermore we introduce a simplified model that 
 allows us to  consider explicitly the case where   the  magnetic configuration evolves in time on timescales comparable to the particle transit time through the configuration. In contrast with previous works on this topic, this analysis  requires that  a  system that is aperiodic in time be investigated. In this case the Poincar\'e map technique cannot be applied  and LCSs remain the only viable tool.}
 
\end{abstract}

\maketitle

\section{Introduction}

The understanding of transport phenomena  in magnetized plasmas is  notoriously one of the most challenging tasks in the  investigation  of  both laboratory and space plasmas.
This is particularly  the case in low collisionality plasmas where particles are essentially free to stream along magnetic field lines while their transport in the perpendicular direction 
is governed by collective electric and magnetic  fluctuations and not by binary interactions.  
As a consequence, transport is generally anisotropic, nonlocal and, most likely, not described by simple diffusion equations. 

In recent years  the  concept  of Lagrangian
Coherent Structures (LCS) has been introduced  by G. Haller, i.e. see Ref. \onlinecite{haller2000lagrangian}, in the context of  transport processes in complex 
fluid flows. In a two-dimensional configuration these structures are now defined, e.g. see Refs. \onlinecite{haller2015lagrangian,haller2011variational}, as special
lines  which are  advected by the fluid  and which organize the  flow. 
The importance of the LCS stems from the fact 
that they {are a generalization of the dynamical structures observed in autonomous and periodic systems, e.g. invariant manifolds, to temporally aperiodic flows, see e.g. Ref. \onlinecite{haller2015lagrangian}.} {Analogously to these structures,} they separate 
the  flow domain into macro-regions inside  which fast
mixing phenomena take place. Over the finite time span which characterizes the LCSs
these macro-regions do not exchange 
fluid elements and thus act as transport barriers.
 The  LCS  concept provides a very  effective tool that is being increasingly used in order  to describe  transport processes in a wide range of conditions:  
the pollutant transport on the ocean surface\cite{coulliette2007optimal}, blood  flow\cite{shadden2008characterization}, the spreading of plankton blooms\cite{huhn2012impact}, jelly fish predator-prey interaction\cite{peng2009transport}, atmospheric dataset
analysis\cite{tang2010lagrangian}, {transport features of the beam-plasma instability\cite{carlevaro2015nonlinear}}, solar photospheric 
flows\cite{chian2014detection}, saturation of a nonlinear dynamo\cite{rempel2013coherent}, etc.

Since  their  introduction, different mathematical definitions of LCS have been proposed with the aim of providing a  tool capable of {giving} a concise representation of the transport processes. The earlier definition involved 
  second derivative ridges of the finite time Lyapunov exponent field, e.g. see  Ref. \onlinecite{shadden2005definition}, but was later corrected and reformulated  in terms  of most repulsive or attractive material lines (see below  in Secs. \ref{LCShal}, \ref{LCSridg}). In addition numerical procedures have been devised (Refs. \onlinecite{farazmand2012computing,onu2015lcs})  in order to extract  such structures  from fluid simulation  results or, even more interestingly,   from actual  experimental data.

In a number of recent articles, Refs. \onlinecite{borgogno2011barriers,rubino2015detection,falessi2015lagrangian}, the description of transport  phenomena in magnetized plasmas has  been addressed  using LCS as a tool  aimed at identifying transport barriers in a  toroidal magnetic configuration  in the presence of  magnetic reconnection events.
Both the  second derivative ridges  and the most repulsive or attractive material lines definitions have been adopted in Ref. \onlinecite{borgogno2011barriers} and in Ref. \onlinecite{falessi2015lagrangian} respectively and results obtained with the two different approaches  have been compared in Ref. \onlinecite{falessi2015lagrangian}.\\
In all these  articles the structure of the magnetic field lines  has been used as a proxy for the structure of the particle trajectories,  assuming that particles move along magnetic field lines  as obtained  from  a ``snapshot''  of the magnetic configuration taken  at a  fixed time, neglecting finite particle orbit size effects, secular drifts 
and assuming that the magnetic configuration does not evolve significantly on the  particle transit time through the configuration.

In the present paper,  {the first one of a linked pair  (called Article {\it I} and  {\it II}  respectively),}  after  briefly  revisiting the ``snapshot''  results, we formulate a  generalization  of the method that, while still not addressing the full particle dynamics, takes nevertheless into account the fact that the magnetic configuration may evolve  on time scales comparable to the particle transit time. 
A consequence of this generalization is that {the system becomes intrinsically aperiodic. Therefore, the Poincar\'e map technique, which has been used in Refs. \onlinecite{borgogno2011barriers,rubino2015detection,falessi2015lagrangian}, cannot be applied to analyze the problem.} We define velocity dependent LCS  i.e. introduce a rudimentary kinetic treatment that  addresses the fact that particles with different energies can be  expected to experience different transport barriers, as has been recently proved by means of test particles simulations in Ref. \onlinecite{borgogno17}.  After these definitions are established,  we will use 
this generalization {in the accompanying Article {\it II}}\,  in order to identify the {LCSs}  in a   magnetic configuration where  magnetic reconnection evolves by referring to the same numerical simulation results that were used in Refs. \onlinecite{borgogno2011barriers,falessi2015lagrangian}.

This paper is organized as follows.\,\, In Sec. \ref{BHam} the Hamiltonian nature of the magnetic field line equation is briefly  rederived in order to  illustrate   the relationship with the dynamics of one-dimensional non autonomous  dynamical systems  having in mind, as will be repeated later in the text, that the ``time'' entering in the 
Hamilton equations for the magnetic field lines is not the physical time  but a  properly chosen coordinate  along   field lines.  A simplified planar magnetic configuration with a strong magnetic field component out of the plane (the so-called guide field) is  considered. The fact that it corresponds to a non autonomous  
dynamical system is  related to the effect of  a process  of magnetic reconnection that has  broken the  underlying structure of magnetic surfaces. This latter configuration would have corresponded to an autonomous (and thus integrable) one dimensional system.  Then the  connection with  particle transport  is recalled and some related early references are mentioned.\\
 In Sec. \ref{Poinc} the distinction between time periodic and aperiodic dynamical systems is  made in connection with  the different mathematical tools that are best suited to describe their dynamics. The role of the  Poincar\'e map for time periodic systems is recalled together with  a short  overview of the so called lobe dynamics. The  Poincar\'e map approach  makes it possible to partition  the phase space of the time periodic dynamical system  into  macro-regions   distinguished by a qualitatively different behaviour of the trajectories they contain, {e.g. periodic or  chaotic  trajectories.}\\
This approach cannot be applied to time aperiodic systems and therefore  in Sec.\ref{LCSnonper}  the concept of  {LCSs} is introduced.  We close  this section by briefly contrasting the initial definitions of LCSs in terms of second derivative ridges, e.g. see Ref. \onlinecite{shadden2005definition}, and the definition in terms of maximal repelling and maximal attracting material lines.

In Sec.\ref{magn}  we first  describe  (Sec.\ref{magnper})   the time periodic dynamical system  related to  the magnetic  configuration  that we choose   (see also Refs. \onlinecite{borgogno2008stable,borgogno2011barriers})  for  the  study of particle transport  in the presence of magnetic reconnection.  Then in Sec.\ref{magnnonper} we introduce  a time nonperiodic dynamical system obtained by including in a simplified way   the effect of the change of the magnetic configuration during  the particle transit  through it.  This is done by combining the  coordinate-like ``time'' entering in the 
Hamilton equations for the magnetic field lines mentioned above with the physical time  that describes the change of the magnetic configuration caused, in our case of interest, by the onset of magnetic reconnection.  The resulting ``effective'' time depends on the velocity of the specific particle  that is considered and can be thought  as the physical time of change of the  local value of the magnetic field seen by the particle along its trajectory because of the combined effect of the magnetic field  spatial inhomogeneity and physical time  evolution.  {The numerical investigation of both the time periodic and the  non-periodic systems will be described in the accompanying paper Article {\it II}.}
\section{Magnetic field as a dynamical system}
\label{BHam}
As is well known \onlinecite{cary1983noncanonical,kruskal1952some,kerst1962influence,gelfand1962magnetic,morozov1966structure,boozer1981plasma}, due to their solenoidal nature, the field lines of a magnetic field in three-dimensional space that does not vanish within the domain of interest  can be described at any fixed physical time $t=\bar{t}$ as trajectories of a non-autonomous Hamiltonian system with one degree of freedom. The role of time is played by a spatial coordinate taken to  label the points  along a  field line.  A simple derivation in terms of a general set of (curvilinear) coordinates $\chi_i$,\, $i = 1,2,3$  can be given by  choosing a gauge condition for the vector potential ${\mathbf A}$ such that  one of its components, e.g. $A_3$, vanishes i.e., 
\begin{eqnarray}
{\mathbf A}\,&=&\,A_{1}\, {\mathbf \nabla}\chi_{1}\,+\,A_{2}\, {\mathbf \nabla} \chi_{2}, \qquad \quad  {\rm and} \nonumber\\  {\mathbf  B}\,&=&\,  {\mathbf \nabla}A_{1}\times {\mathbf \nabla}\chi_{1}\,
+\,  {\mathbf \nabla}A_{2}\times {\mathbf \nabla}\chi_{2}.
\end{eqnarray}
Since 
$\ {\mathbf B}\neq 0$ within the considered spatial domain  we can set
$\ {\mathbf B}\cdot\ {\mathbf \nabla}\chi_{1}\,\neq\,0$. 
It follows that the Jacobian 
$\left( {\mathbf \nabla}A_{2}
\times {\mathbf \nabla}\chi_{2}\right)\cdot{\mathbf \nabla}\chi_{1}\,\neq\,0$,
i.e. that  the coordinate transformation to the new set of  spatial coordinates $A_{2}, \chi_{2},\chi_{1}$ is invertible. \\Using $ {\mathbf B}\,=\,
 {\mathbf\nabla}A_{1}({\chi}_{1},{\chi}_{2},A_2)
\times {\mathbf \nabla}{\chi}_{1}\,
+\, {\mathbf \nabla}A_2\times {\mathbf \nabla}{\chi}_{2} $,
from the field line condition  $d{\mathbf l}\times\ {\mathbf B}=0 $ we obtain the Hamilton equations 
\begin{eqnarray}\label{gen}\nonumber
&& \frac{dA_{2}}{d
\chi_{1}}
\, = \,   \frac{\partial A_{1}}{\partial\chi_{2}}, \quad 
\frac{d\chi_{2}}{d\chi_{1}}
\,=\,-\frac{\partial A_{1}}{\partial A_{2}}\\ && {\rm together \, ~\, with } \qquad  \frac{d A_{1}}{d\chi_{1}}
\, =  \, \frac{\partial A_{1}}{\partial \chi_{1}}. 
\end{eqnarray}
In the above equations  $\chi_{1}\,$  plays the role of 
the ``time'' variable,  $\chi_{2}$ and $A_{2}\,$ that of the two canonical coordinates:  $d A_{2}\leftrightarrow dq$ and $d \chi_{2} \leftrightarrow dp$, and  $A_{1}(\chi_2,A_2, \chi_1)$
of  the ``Hamiltonian'' $\mathcal{H}$. 
We anticipate here that in the following we will consider  a simple magnetic field configuration  often used in the study of magnetically confined plasmas where   $  {\mathbf B} \,=\,  B_z  {\mathbf e}_z + {\mathbf \nabla}\psi\times\ {\mathbf e}_z$, with $B_z$ spatially uniform and $  {\mathbf e}_z = {\mathbf \nabla}z$.  It corresponds,  after an appropriate rescaling, to
{$\chi_{1} =   z,$\, $\chi_{2} = x $, \, $A_{2} = y$ and $A_{1}(x,y,z)$}
the so-called ``poloidal'' flux function $\psi(x,y,z, t=\bar{t}) $.

The importance of this Hamiltonian formulation stems from the fact  that  it establishes a direct connection between magnetic configurations and dynamical systems, see e.g. Refs. \onlinecite{rosen1,elsasser,morrison,boozer2005physics},        
and  thus makes it possible to describe the topology of the magnetic field lines in terms of that of the trajectories of a  dynamical system.  Furthermore, if we assume that  in the considered magnetic configuration an adiabatic approximation holds for the  motion of the charged plasma particles, in the limit where their Larmor radius is negligible and disregarding  the particle drifts,  we can approximate  their motion as occurring  along magnetic field lines.  
Such an approximation allows us to study the particle advection  and diffusion, see e.g. Ref. \onlinecite{rosen2},  by using the same set of equations that determine the magnetic field lines.  This  approach requires in addition that,  as a first approximation,    the change of the magnetic configuration during the particle motion be neglected.

The equivalence between the magnetic field lines and the trajectories of non-autonomous Hamiltonian systems with one degree of freedom has been  widely  used in the literature by adopting the concepts that are  proper of dynamical systems, see in particular Ref. \onlinecite{cary1983noncanonical}. In the case of magnetically confined plasmas this equivalence has been used in particular  in order to asses  the effects of magnetic field lines reconnection events on the particle transport.
In the context of the present article we will refer to Ref. \onlinecite{borgogno2008stable} and more specifically  to Refs. \onlinecite{borgogno2011barriers,falessi2015lagrangian} where dynamical systems tools are used in a systematic way in order  to characterize transport processes associated to the magnetic field lines topology. This  approach allows us to partition the magnetic configuration  into sub-domains characterized by different {transport phenomena} and in particular  to identify the domains where {they are} either fastest or slowest.

\section{Transport phenomena in non-autonomous, time periodic, dynamical systems with one degree of freedom} \label{Poinc}

Non-autonomous  dynamical systems may have a periodic or non-periodic time dependence, see e.g. Refs. \onlinecite{arnold2007mathematical,wiggins1992chaotic,cencini2010simple}, and the techniques used to study these two cases are very different.  In particular, when dealing with a periodic system the  Poincar\'e section (stroboscopic map) method can be used to reduce the dimensionality  of the problem by  studying a $2 N$ dimensional map instead of a $2N +1$ continuous-time dynamical system. This map is constructed by associating to an initial condition its evolution after one period. The main advantage obtained by using this technique is to filter out redundant dynamical phenomena and reveal the underlying nature of the motion, e.g. whether it  is regular or chaotic. Furthermore, {invariant} curves of the  Poincar\'e map can be used to partition the phase space into regions where trajectories have a qualitatively different behaviour  {on a given time-scale}, e.g. bounded or unbounded\cite{mackay1987resonances,meiss2015thirty,ottino1989kinematics}. These structures play a fundamental role in governing transport processes in non-autonomous dynamical systems and, in particular, they determine the so-called lobe dynamics. 
In this section  we will briefly recall the main features of lobe dynamics while extensive presentations can be found in Refs. \onlinecite{wiggins2013chaotic,rom1990transport,rom1990analytical,malhotra1998geometric}. 

\begin{figure}
\centerline{\includegraphics[width=9cm]{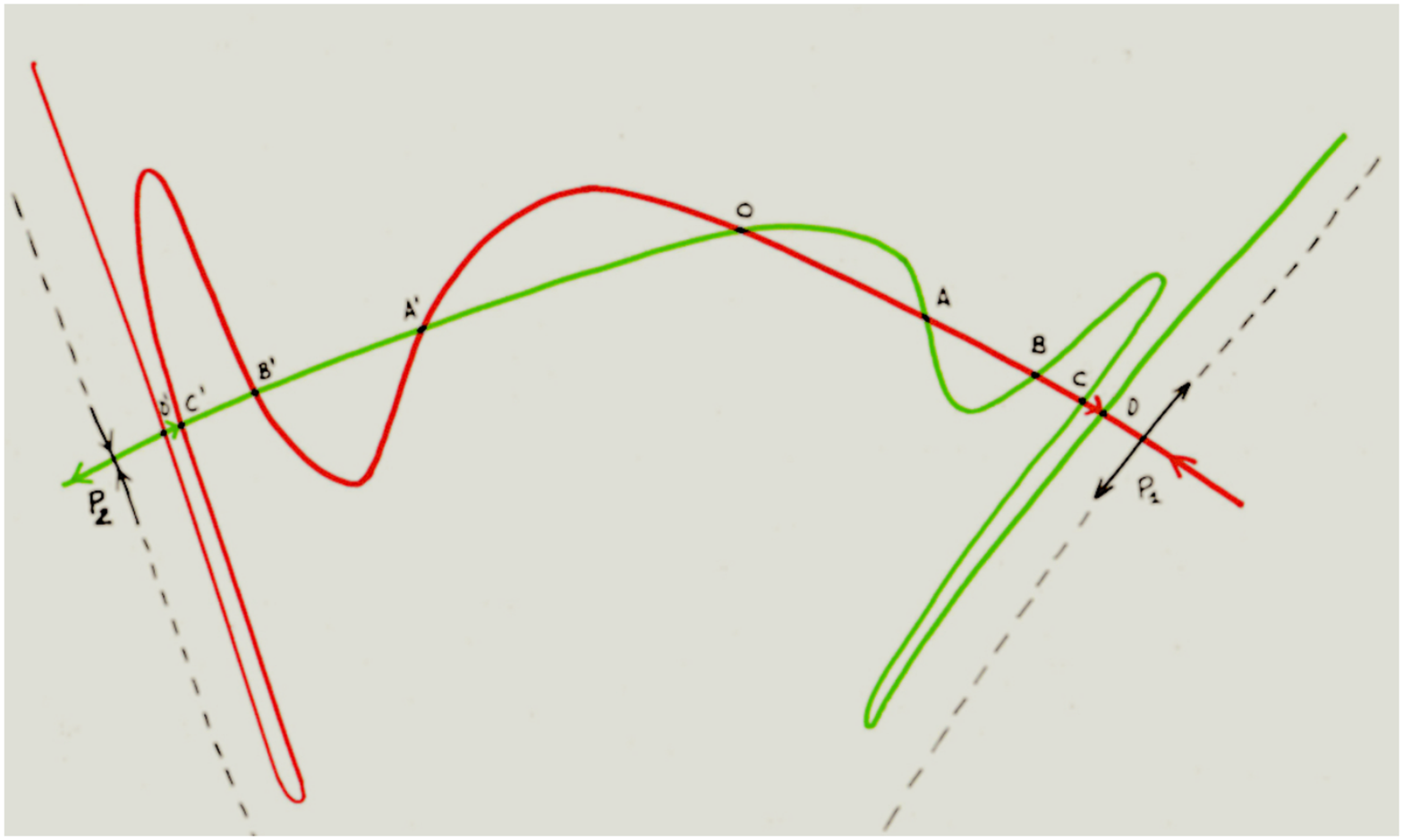}}
 \caption{Sketch of the    lobes  produced by the convoluted shape of  the stable and  the unstable manifolds in the proximity of two DHTs.  }
\label{fig1}\end{figure}

In  view of the present application to the study of the topology of magnetic field lines at fixed physical time,  we consider   explicitly systems with one degree of freedom (N=1) and, in order  to avoid any misunderstanding, we recall that the role of  the  independent variable of the Hamiltonian system in the discussion that follows is played not by  the physical time but by the $\chi_1 = z$ coordinate, as mentioned  above. Following Ref.
 \onlinecite{malhotra1998geometric}, we define a lobe 
 as a region of  the {\it extended} ($2N+ 1$ dimensional) phase space enclosed by segments of the intersection between stable and unstable manifolds and a  $t=const$ surface (i.e., here, at a  $z=const$ surface, this remark will not be repeated in the rest of the section and in the following ones until the true physical time is reintroduced in Sec. \ref{magnnonper}). 
We recall that stable and unstable manifolds are defined with respect to a distinguished hyperbolic trajectory (DHT) i.e. with respect to a special trajectory that shares the property  of being a solution  of the non-autonomous Hamiltonian equations and of connecting instantaneous (i.e. at fixed time) hyperbolic points (so called $X$-points), see e.g. Refs. \onlinecite{wwiggins1992chaotic,borgogno2008stable}. It  has the property that all neighbouring field lines approach such a  trajectory exponentially either forward or backward in time.
Stable and unstable manifolds  are invariant surfaces defined  as the set of trajectories  that converge towards
the DHT forward or backward in time, respectively. The intersection of these  manifolds with a  $t=const$ surface 
define one dimensional curves. A sketch of these curves and of the  lobes  produced by the convoluted shape of  the stable and  the unstable manifolds in the proximity of two DHTs is shown in Fig.\ref{fig1}. 
 \\ From the definition of lobes it follows that their geometry evolves in time and that in the particular case of time periodic systems it must remain  constant on each  Poincar\'e section. Since stable and unstable manifolds are invariant sets, the points belonging to the intersection of these two manifolds must be mapped into points which belong to their intersection at each iteration of the map. These two statements, together with the orientation {\it of the arrows} in the figure of the stable and of the unstable  intersecting manifolds, which  indicates  a clockwise motion, allows us to follow  the lobe motion at each  iteration of the  Poincar\'e map. Let us  start by considering the lobe enclosed by the two segments $OA$ where $O$ is a chosen, but otherwise arbitrary, intersection point.  After one iteration of the  Poincar\'e map  the segments $OA$ and the enclosed  lobe move to $BC$. Due to the Hamiltonian nature of the system the area of the  lobe must remain constant and therefore the lobe stretches in order to compensate the compression in the direction parallel to the stable manifold of $P_{1}$. In summary the lobe  moves  clockwise and, at the same time, is deformed into a thin structure. Since we are interested in studying transport processes it is useful, having selected the intersection point $O$, to split  phase space into two regions separated by the boundary $P_{2}O P_{1}$ taken along the unstable manifold of $P_2$ and the stable manifold of $P_1$, see e.g. Refs. \onlinecite{rom1990transport,rom1990analytical}. 
This makes it possible  to split the phase space into two sub-regions with a relatively simple boundary with respect to other intersection points. Considering this partition, after one iteration of the map, the lobe $O A$ remains in the region  from where it started its motion. This is  the case for the other lobes  but it is not the case for the lobe $B' A'$,  which is mapped by the dynamics into $O A$ and it is therefore ejected from the region from which it started, and for  the lobe $O A'$ which starts inside the upper region and  is mapped  to  $A B$ into  the lower region. For this reason the two lobes, i.e. $O A'$ and $B' A'$, are often called turnstiles, see e.g. Refs. \onlinecite{wiggins1992chaotic,rom1990transport,rom1990analytical}. This dynamics shows that transport process  across the boundary $P_{2}O P_{1}$ consists in the exchange of one lobe per iteration of the map. Any other initial condition remains in the region where it started.
Therefore phase space turns out to be partitioned into macro-regions, see Refs. \onlinecite{mackay1987resonances,chen1987area,raynal2006lobe}, where trajectories have a qualitatively different behaviour with respect to the trajectories belonging to the other ones.
How fast mixing phenomena develop depends on the shape of the  lobes. In fact, if the  area of the lobes is sufficiently small, for most of the initial conditions the trajectories get stirred by the dynamics, while transport  across the boundary $P_{2}O P_{1}$  consists in the exchange of relatively small phase space structures, i.e. the lobes.

\section{Transport phenomena in non-autonomous, non periodic, dynamical systems with one degree of freedom} \label{LCSnonper}

In the previous section we have briefly recalled how it is possible to characterize transport processes in periodic systems using the geometry  of  stable and unstable manifolds. These structures  can be calculated knowing the trajectories and the velocity field of the dynamical system 
 for the  finite time interval given by the system periodicity.
 
This is not the case of non periodic-systems. In fact, in the most general case,  knowledge of the velocity field is required for an infinite time interval in order to define invariant manifolds and lobes. For this reason a different technique,  based on the definition and identification  of  the so-called Lagrangian Coherent Structures (LCS), e.g see Refs. \onlinecite{haller2015lagrangian,haller2011variational,shadden2005definition},  has been developed in order to study transport processes for the most general non periodic flows.

As in the periodic case, the aim is to identify domains in phase space with different dynamical behaviour and, eventually, transport features. The LCS are the boundaries of these \emph{regions of coherence}. In contrast to  the invariant manifolds, LCS are defined over a finite amount of time, $\tau$,  related to the characteristic time of coherence of the motion. As $\tau$ increases  LCS converge to the invariant manifolds mentioned above,  see Ref. \onlinecite{haller2015lagrangian}. We note in passing that, although  not directly relevant to our problem, the concept of LCS does not require the system to be Hamiltonian.

Following Refs. \onlinecite{falessi2015lagrangian,borgogno2011barriers}, in this work we consider only hyperbolic LCS which organize the Hamiltonian flow  by attracting or repelling volume elements of  phase space  over the finite time span $\tau$. For the sake of simplicity we will refer to these structures simply as LCS. The rigorous definition of these structures has been subject to debate: see e.g. Refs. \onlinecite{shadden2005definition,haller2011variational,falessi2015lagrangian}. A first way of finding LCS was based on constructing  the  field of the finite time Lyapunov exponents  (FTLE), see Refs. \onlinecite{shadden2005definition,borgogno2011barriers,rubino2015detection}. Since  at a given phase space position the largest positive FTLE measures the exponential separation between two neighbouring  initial conditions after a given interval of time, within this formulation LCS have been defined as second-derivative ridges of the  FTLE-field. Several counter examples to this heuristic definition have been found by G.Haller who introduced the 
definition of hyperbolic LCS, see Ref. \onlinecite{haller2011variational}, as the most repulsive or attractive material lines, where material lines are defined as  lines of initial conditions advected by the Hamiltonian flow.

\section{Lagrangian Coherent Structures (LCS)} \label{LCgen}

In this Section we briefly  recall the   definition  of LCSs  (see  \onlinecite{haller2011variational,falessi2015lagrangian}). 
As mentioned before, we consider a dynamical system  in 2D phase space ${\boldsymbol  x} = (x,y)$,
\begin{equation}\label{1=}
\frac{d{x}}{dt}={v_x}(t,{x,y}),\qquad \frac{d{y}}{dt}={v_y}(t,{x,y})
\end{equation}
with  continuous differentiable flow map 
\begin{equation}\label{2=}
 {\boldsymbol \phi}_{t_0}^t({\boldsymbol  x}_0)={\boldsymbol  x}(t,t_0,{\boldsymbol  x}_0).
\end{equation}
Two neighbouring points ${\boldsymbol x}_0$ and ${\boldsymbol x}_0+ {\boldsymbol   \delta  x}_0$ evolve into the points ${\boldsymbol x}$ and ${\boldsymbol x}+ {\boldsymbol \delta x}$ under the linearized map
\begin{equation}\label{3=}
{\boldsymbol  \delta x} ={\boldsymbol   \nabla} {\boldsymbol \phi} _{t_0}^t \,  {\boldsymbol\delta   x}_0.
\end{equation}
Let us consider a curve $\gamma_0=\{{\boldsymbol x}_0=r(s)\}$ and 
at  each point  ${\boldsymbol  x}_0 \in \gamma_0$ let us define the unit tangent vector ${\boldsymbol  e}_0$ and the normal vector ${\boldsymbol  n}_0$. In the  time interval $[t_0,t]$ the dynamics of the system advects the {\it material line} 
$\gamma_0$ into $\gamma_t$ and ${\boldsymbol  x}_0\in \gamma_0$ into ${\boldsymbol  x}_t \in \gamma_t$. The linearized dynamics maps the tangent vector ${\boldsymbol  e}_0$ into ${\boldsymbol  e}_t$ which is tangent to $\gamma_t$  
and is given by 
\begin{equation}
{\boldsymbol  e}_t =
\frac{\boldsymbol  {\nabla} {\boldsymbol \phi}^t_{t_0}({\boldsymbol  x}_0)\, {\boldsymbol  e}_0}{ [ {\boldsymbol  e}_0\,  {\boldsymbol  C}^t_{t_0}({\boldsymbol  x}_0) \, {\boldsymbol  e}_0]^{1/2}},
\label{tangent}
\end{equation}
where ${\boldsymbol  C}^t_{t_0}({\boldsymbol  x}_0)\equiv \left(\boldsymbol  {\nabla} {\boldsymbol \phi}^t_{t_0}\right)^T \, \boldsymbol  {\nabla} {\boldsymbol \phi}^t_{t_0}$  is the \emph{Cauchy-Green strain tensor} and $^T$ stands for transposed. This symmetric tensor describes the deformation of an arbitrarily small circle of initial conditions, centered in ${\boldsymbol  x}_0$ caused by the  flow  in a time interval $[t_0,t]$. Taking for example  a circle centered in ${\boldsymbol  x}_0$ with radius $\|\delta {\boldsymbol  x}_0\|$, after the time interval $[t_0,t]$ it will be deformed into an ellipse with major axis in the direction of ${\boldsymbol  \xi}_{max} $ and minor axis in the direction of ${\boldsymbol  \xi}_{min} $, where  ${\boldsymbol  \xi}_{max}$ and ${\boldsymbol  \xi}_{min}$ are the two eigenvectors of ${\boldsymbol  C}^{t}_{t_0}({\boldsymbol  x}_0)$. 
The corresponding real and positive eigenvalues are $\lambda_{max}$ and $\lambda_{min}$. 
The curves with tangent vector along  ${\boldsymbol  \xi}_{min} $  and, respectively,  ${\boldsymbol  \xi}_{max} $ are called \emph{strain lines} of the Cauchy-Green tensor.
In general the mapping  does not preserve the angle between vectors and therefore  usually $ {\boldsymbol  n}_t$  differs from ${\boldsymbol  \nabla} {\boldsymbol \phi}^t_{t_0}  {\boldsymbol n}_0$. 

Using the orthogonality condition  ${\boldsymbol  n}_0 \cdot {\boldsymbol  e}_0  ={\boldsymbol  n}_0 {\boldsymbol  \nabla}{\boldsymbol \phi}^{t_0}_t{\boldsymbol  \nabla}{\boldsymbol \phi}^{t}_{t_0}{\boldsymbol  e}_0=0$ and inserting  Eq. (\ref{tangent}) we obtain the expression for ${\boldsymbol  n}_t$ which is given by 
\begin{equation}
{\boldsymbol  n}_t =
 \frac{\left(\boldsymbol  {\nabla} {\boldsymbol \phi}^{t_0}_{t}\right)^T{\boldsymbol  n}_0}{ [ {\boldsymbol  n}_0\,  {\boldsymbol  C}^{-1}({\boldsymbol  x}_0) \, {\boldsymbol  n}_0]^{1/2}}, 
\end{equation}
where ${\boldsymbol  C}^{-1} ({\boldsymbol  x}_0 ) = {\boldsymbol  C}^{t_0}_t ({\boldsymbol  x}_0 ) $ and the time interval marks have been  suppressed  as will be the case in the following formulae when not explicitly needed.

We  define the \emph{repulsion ratio} $\rho^t_{t_0}({\boldsymbol  x}_0,{\boldsymbol  n}_0)$ as the ratio at which material points, in other words points advected by the flow, initially taken  near the point ${\boldsymbol  x}_0 \in \gamma_0$,   increase  their distance from the curve in the time interval $[t_o,t]$:   
\begin{equation}
\rho^t_{t_0}({\boldsymbol  x}_0,{\boldsymbol  n}_0)={\boldsymbol  n}_t \,  {\boldsymbol \nabla} {\boldsymbol \phi}^t_{t_0}({\boldsymbol  x}_0) \, {\boldsymbol  n}_0.
\label{repulsion-ratio}
\end{equation}
Using the previous definitions, $\rho^t_{t_0}({\boldsymbol  x}_0,n_0)$ can be expressed either in terms of $n_0$ or of  $n_t$ as 
\begin{equation}
\rho^t_{t_0}({\boldsymbol  x}_0,n_0)= [ {\boldsymbol  n}_0\,  {\boldsymbol  C}^{-1}({\boldsymbol  x}_0) \, {\boldsymbol  n}_0]^{-1/2}=  [ {\boldsymbol  n}_t\,  {\boldsymbol  C}({\boldsymbol  x}_0) \, {\boldsymbol  n}_t]^{1/2} 
\end{equation}
Similarly, the \emph{contraction rate} $L^{t}_{t_0}({\boldsymbol  x}_0)$ is proportional to the growth in time of the vector tangent to the material line
\begin{equation}
L({\boldsymbol  x}_0, {\boldsymbol  e}_0)= [ {\boldsymbol  e}_0\,  {\boldsymbol  C}({\boldsymbol  x}_0) \, {\boldsymbol  e}_0]^{1/2} .
\end{equation} 

\subsection {LCS as maximal repulsion-attraction material lines}  \label{LCShal}

Here we  adopt the definition of a Hyperbolic LCS as given in Ref.\onlinecite{haller2011variational}. \, An LCS over a finite time
 interval $\left[t_0, t_0+T \right]$ is defined as a material  line along which the repulsion rate is pointwise maximal.
This leads, as shown   in  Refs. \onlinecite{haller2011variational,falessi2015lagrangian}, to the following definitions.\\
A material line satisfying the following conditions at each point:
 \begin{equation} a) \qquad \quad  \lambda_{min}<\lambda_{max},\quad \lambda_{max}>1 , \label{condizionelambda}\end{equation}
\begin{equation} b) \qquad  \qquad   \qquad  
	{\boldsymbol e}_0=  {\boldsymbol \xi}_{min}  \qquad   \qquad 
\label{condizione2}
\end{equation}
the tangent vector is along the eigenvector associated with the smallest eigenvalue,
\begin{equation} c) \qquad  \quad   \quad   \,
{\boldsymbol  \xi}_{max} \cdot {\boldsymbol  \nabla} \lambda_{max} =0 \qquad  
\label{numericallyaproblem}
\end{equation}
 the gradient of the largest eigenvalue is along the curve, is called a {repulsive} Weak Lagrangian Coherent Structure (WLCS).
\\ A WLCS which satisfies   at each point the additional condition
\begin{equation}\label{maxrep}
{\boldsymbol  \xi}_{max} \, \cdot  {\boldsymbol   \nabla }^2 \lambda_{max}\,  \cdot \, {\boldsymbol   \xi}_{max}  <0
\end{equation}
 is called a {repulsive} Lagrangian coherent structure. {Attractive LCS are defined as repulsive LCS of the backward time dynamics}.
 
Finally we note that  in the case of a Hamiltonian non-autonomous system with one degree of freedom phase space conservation implies
\begin{equation}
 \lambda_{min}\,  \lambda_{max} = 1.
\end{equation}

 \subsection {LCS as second derivative ridges}\label{LCSridg}

 An alternative definition of LCS  was proposed in Ref. \onlinecite{shadden2005definition}   in terms of 
second derivative ridges. These ridges were introduced in order  to describe  the  essential features   of the finite time  Lyapunov exponent field $\sigma ( {\boldsymbol  x}_0, t_0, t )$ that 
characterizes the rate of separation in the time interval $ t_0, t $ of close trajectories of the flow. We recall that the  finite time  Lyapunov exponents  can be expressed  in terms of the   eigenvalues  of the Cauchy-Green tensor as 
 \begin{equation}
 \label{Lyap}
 \sigma ( {\boldsymbol  x}_0, t_0, t ) = \frac{1}{ 2|t -t_0|}\, \ln{\lambda_{max}} ( {\boldsymbol  x}_0, t_0, t ) 
 \end{equation} 
Second derivative ridges (SDR)  are defined  in terms of the  Hessian matrix ${\boldsymbol  \Sigma }$  of the second derivatives of $\sigma$ with respect to $ {\boldsymbol  x}_0$.
SDR are curves  ${\boldsymbol r}(s)$ (not necessarily   material lines) for which 
 the tangent vector  ${\boldsymbol r}^\prime(s)$ and  $\,  {\boldsymbol \nabla}\sigma$ along the curve are parallel,   $\, \,  {\boldsymbol  u} \,   {\boldsymbol    \Sigma} \, {\boldsymbol  u}   < 0 \,  $ for all  unit vectors ${\boldsymbol u}$
and   the normal unit vector $ {\boldsymbol  n}$ is such that   \begin{equation}
 \label{Sigma}
  {\boldsymbol  n} \,   {\boldsymbol    \Sigma} \, {\boldsymbol  n}  = {\rm min} \end{equation}
  when the Hessian$\ {\boldsymbol  \Sigma }$ is  evaluated at the curve. 
  
  This definition was criticized both on mathematical grounds, as being inconsistent  under generic conditions, see Ref. \onlinecite{haller2011variational},  and on physics grounds since the curves it identifies are 
  not  material lines and thus  the flow through them need not vanish (or even in specifically constructed counter-examples be small).
A major difference  is that the definition of LCS as maximal repulsion-attraction material lines involves the eigenvectors and eigenvalues of the Cauchy-Green strain tensor ${\boldsymbol  C}$
while the definition in terms of  the second derivative ridge is governed by
the eigenvectors and eigenvalues of the Hessian  ${\boldsymbol  \Sigma }$. In Ref. \onlinecite{falessi2015lagrangian} an elementary  example is  discussed where a WLCS is explicitly  shown not to be  a second derivative ridge.

\section{Reconnecting magnetic configuration} \label{magn}

As mentioned in the Introduction,  { the  aim of this  and its accompanying paper } is to obtain information about particle transport due to the onset of magnetic reconnection from the behaviour of {the} magnetic field lines.  
The reconnection setting that we adopt is the same as that used  in  Ref. \onlinecite{borgogno2011barriers},
which is  based on a  numerical simulation reported in  Ref. \onlinecite{borgogno2005aspects}, where reconnection is made possible by the effect of electron inertia. The reconnecting
magnetic field has only components in the $x$-$y$ plane but depends on all the three spatial
coordinates.  In this numerical  simulation
the magnetic field evolution starts from a static equilibrium, expressed in terms of a magnetic flux function $\psi_{eq}$ as
\beq\label{equil}
    {\bf B}_{eq} =  B_0{\bf e}_z + \nabla\psi_{eq}(x)\times {\bf e}_z ,
\eeq
with
$\psi_{eq} = 0.19\cos(x)$. Periodicity is assumed in all three directions and the configuration is restricted to the domain 
$[-L_x,L_x]\times [-L_y,L_y]\times [-L_z ,L_z]$  with   $L_x = \pi, L_y = 2\pi, L_z =16\pi$.
\\
A ``double helicity'' perturbation (i.e., in the considered planar geometry,  a perturbation made of two components with different phase planes) is initially imposed 
\begin{eqnarray}\label{pert}
\hat{\psi}(x, y, z, t) =  &&\hat{\psi}_1(x, t) \cos{(k_{1y}y + k_{1z}z)} + \nonumber \\ && \hat{\psi}_2(x, t) \cos{(k_{2y}y + k_{2z}z)} ,
\end{eqnarray}
where  $k_{1y} = k_{2y}  =  2\pi/L_y$ and $k_{1z} = 0$ while  $k_{2z} = 2\pi/L_z$. The
amplitude of $\hat\psi_1$  was  chosen to be of order $10^{-4}$ and  ten times bigger than that of $\hat{\psi_2}$.
As is well known,  perturbations with different ``helicities''   are required in order to make the  Hamiltonian system  described in Sec. \ref{BHam}  non integrable, i.e., to generate a chaotic magnetic configuration.
In the  following we will denote by  $\psi(x,y,z,t) = \psi_{eq} (x) + \hat{\psi}(x, y, z, t) $ the complete magnetic flux function that includes the equilibrium and the physical time evolving perturbations. At each fixed physical time $t$ 
it plays the role of the Hamiltonian for the magnetic field lines while the space coordinates  $x$  and $y$  that
of   canonical variables  with  $x$ the momentum and $y$ the position.
The field line Eq.(\ref{gen}) becomes
\begin{equation}  \label{psin1} \frac{d x}{d z} = - \frac{\partial  \psi}{\partial y} , \qquad  \frac{d y}{d z} = \frac{\partial  \psi}{\partial x} . \end{equation}
In the linear phase, when the two components of  the perturbation evolve  independently without interacting with  each other, two chains  of magnetic islands are  formed   around their own resonant surfaces.  These  are defined by the condition ${\bf B}_{eq}\cdot {\bf k}_{1,2}=0$ and, disregarding the  mirror-doubling of the configuration  caused by the  assumed periodicity along $x$, are located at $x_{1} = 0$ and
$x_{2} = 0.71$ respectively.
During their evolution the magnetic islands expand and  start to interact making the linear approximation invalid. The dynamics of the magnetic configuration  becomes rapidly nonlinear  and higher  order modes are {spontaneously} generated. The most relevant of these nonlinear modes turn out to  have   the same  helicities of the two components of the imposed perturbation. At this stage, the magnetic field topology exhibits regions where field lines are stochastic and these regions tend to spread as  the reconnection process evolves. A detailed description of the chaos inception and spread all over the  domain of the configuration can be found  in Refs. \onlinecite{borgogno2005aspects,borgogno2008stable}.

In the { numerical investigation in Article {\it II}}  we will focus on two different normalized  physical  times, $t = 415$ and  $t = 425$ Ref.  \onlinecite{borgogno2008stable} (normalized respect to the Alfv\'en time defined in terms of $\psi_{eq}$), in which the chaos, initially developed only  on a local scale (at $t=415$), starts to  spread on a global scale (at $t=425$).

\subsection{Time periodic dynamical system} \label{magnper}

First we consider the dynamical system that is obtained by taking  a snapshot   at a given physical time $t = \bar{t}$ of the reconnecting  magnetic configuration  following  the procedure introduced in Sec. \ref{BHam}, {where  the flux  function $\psi(x,y,z,t = \bar{t})$ is the Hamiltonian
and  $z$ is the magnetic Hamiltonian  time}. Since the configuration is periodic in $z$  with periodicity $32 \pi$  we can adopt   the  Poincar\'e map technique  and compare it with the LCS approach. 
The result of this comparison  is shown in Sec. \ref{resmagnper} of Article {\it II}  for one snapshot  taken respectively at  $t =415$ .
\\
The  magnetic configuration in Sec.\ref{magn} is symmetric under the space-time reflection symmetry $y\to -y, \,\, z\to -z$ since 
$\psi(x,y,z,t= {\bar t}) = \psi(x,-y,-z,t= {\bar t})$. 
This property can be exploited when computing attractive LCSs as they can be seen  as repulsive LCSs  with respect to the  inverted ``time'' $-z$.
Because of the  above reflection symmetry this time inversion is equivalent to  setting  $y\to -y$, i.e., the attractive LCSs are mirror images of the repulsive LCSs with respect to $y=0$.

\subsection{Time nonperiodic dynamical system }  \label{magnnonper}

A rudimentary way to take into account  the fact that the  magnetic configuration changes during the particle transit time  is to adopt a model where the particle gyromotion and  drifts  are neglected and the particles dynamics 
is only included  through their  streaming  velocity $V$ along the  guide field  $B_0$ i.e., along $z$.  Furthermore $V$  is  assumed to stay  constant.  As mentioned in the Introduction, this model  oversimplifies the description of the  particle  transport  caused  to the onset of magnetic reconnection.
However it  allows us to  describe LCSs in a time nonperiodic dynamical system and, most importantly, to include kinetic-type effects by  defining LCSs that depend explicitly on the different particle velocities.

With this in mind, we introduce a  family of  nonautonomous dynamical systems   in the extended phase space $x,y,z$, with $z$ playing again the role of time and  each system  being characterized by a different velocity $V$, 
by introducing the Hamiltonian
\begin{equation}\label{psiV}
\psi_{\cal V}(x,y,z)\,\, \equiv \,\,  \psi(x,y,z, t = (z-z_o)/V). \end{equation}
Here $t$ is taken to be positive and, in fact,  it is defined at $t-t_1$ where $t_1$ is the  physical  time  at which we start our investigation of the particle trajectories while, 
 for convenience, we set  the starting  magnetic Hamiltonian time $z_o = 0$. In the following we will restrict the range over which the physical time $t$ varies to the interval 
 $t_1 < t < t_2$.  As mentioned above, at  $t_1 = 415$,  i.e. in the  early stage of the nonlinear
reconnection process, the regions of chaoticity of the magnetic field lines are still separated. At the  later stage $t_2 = 425$ {chaotic regions merge and the system experience a transition from local to global chaoticity}.

\subsubsection{Positive  and  negative  velocities } \label{+-}
For positive velocities  the physical time $ t = z/V $  that appears in the Hamiltonian $ \psi(x,y,z, t = (z-z_o)/V)$ increases as $z$ increases
and the new Hamilton equations read
\begin{equation}  \label{psiV1} \frac{d x}{d z} =- \frac{\partial  \psi_{\cal V}}{\partial y} , \qquad  \frac{d y}{d z} =  \frac{\partial  \psi_{\cal V}}{\partial x} . \end{equation}
On the contrary for negative velocities $ z$ decreases as $t $ increases. Thus for negative velocities  it is convenient to refer to the variables  $\zeta = - z$ and $\eta = - y$  and write \\
\begin{eqnarray}\label{psi-V}  \psi_{-|{\cal V}|}(x,y,z) \, & \equiv&  \, 
\psi(x,y,z, t = -z/|{ V}|) \nonumber  \\ &=& \,  \psi(x,\eta,\zeta, t = \zeta/|{ V}|), \end{eqnarray}
where we have used the symmetry at fixed  time $t$, so that 
$ \psi_{-|{\cal V}|}(x,y,z) \equiv \psi_{|{\cal V}|} (x,\eta,\zeta)$.\\
Then for negative velocities the Hamilton equations (\ref{psi-V})
can be rewritten in the form
\begin{equation}\label{psi-V1}  \frac{d x}{d \zeta} = -\frac{\partial  \psi_{|{\cal V}|}}{\partial \eta } ,  \qquad  \frac{d \eta}{d \zeta} =\frac{\partial  \psi_{|{\cal V}|}}{\partial x}.\end{equation}
which shows that the trajectories for positive and negative velocities differ since they are determined by the same Hamiltonian but  involve different spatial and time domains.

\subsubsection{Repulsive and attractive velocity dependent LCS} \label{Repattr}

As shown above, the time periodic Hamiltonian  attractive LCSs are simply mirror images of the repulsive LCSs with respect to $y = 0$. 
In the case of  the time nonperiodic Hamiltonian  this  result is no longer valid. \\ Here, for the sake of  clarity  we consider only positive velocities $V$.
For attractive LCS we start from $t = t_2$  and $z_{fin} = z+  (t_2 -t) V$ and find it convenient to define a new Hamiltonian function  (with a reversed sign because of the inversion of the time variable $z$)
\begin{equation}\label{psi-attr} {\bar \psi} (x,y,  z_{fin}-z, t_2-t)  = -   \psi (x,y,  z, t),  \end{equation}
and  the variable ${\bar \zeta} = z_{fin}-z,$ (shifted with respect to the variable $\zeta$  defined  in Sec.\ref{+-}). At fixed physical time $t$ we have 
\begin{equation}\label{psi-attr1} \frac{d x}{d z} = -\frac{\partial  \psi }{\partial y} \rightarrow  \frac{d x}{d{\bar  \zeta}} = -\frac{\partial  {\bar \psi}}{\partial y} , \quad \frac{d y}{d z} = \frac{\partial  \psi }{\partial x} \rightarrow  \frac{d y}{d {\bar \zeta} }= \frac{\partial  {\bar \psi}}{\partial x} . \end{equation}
Proceeding as for Eq.(\ref{psiV}) we 
 write
\begin{equation}\label{psi-attr2}   {\bar \psi}_{\cal V} (x,y,  {\bar \zeta} ) =   {\bar \psi} (x,y,  {\bar \zeta},{\bar \zeta}/V)  , \end{equation}
and obtain
\begin{equation}\label{psi-attr3}  \frac{d x}{d {\bar \zeta}} = - \frac{\partial  {\bar \psi}_{\cal V}}{\partial y} , \qquad
\frac{d y}{d{\bar \zeta}} = \frac{\partial  {\bar \psi}_{\cal V}}{\partial x} ,  \end{equation}
which shows that  the equations for the attractive LCS  are the same  in form as those for the repulsive LCS  but with  involve a different  ``time'' variable, ${\bar \zeta}$,  and a different Hamiltonian, $ {\bar \psi}_V$.

{\section{Conclusions} \label{concl1}

In the first  (Article {\it I}) of a pair of linked papers, we have presented the theoretical framework that  will be used in Article {\it II} \, for  the numerical investigation of the Lagrangian Coherent Structures  (LCS) seen by plasma particles restricted  to move along the magnetic field lines of a magnetic configuration that evolves in physical time because of magnetic reconnection. Our  aim is to identify   macro-regions   distinguished by a qualitatively different behaviour  of the particle motion.
The main  limitation of this simplified model  arises from this restriction on the particle motion  that,  however,  it plays a very convenient role as it  allows us to extend the scope of the well known 
relationship  between field line equations and the dynamics of a non autonomous dynamical system with one degree of freedom.

Clearly this restriction can be overcome by inserting into Eqs. (\ref{1=})    a more realistic expression for  the particle motion,  as obtained e.g.  in the gyrocenter approximation} {(see Refs. \onlinecite{frieman1982nonlinear,brizard2007foundations}),} {once the magnetic  and electric  field configurations and  their  time evolution are known.
In order to maintain a description that is two-dimensional in space  plus time,  as in  the simplified treatment described above, the particle trajectories need be expressed with respect to  a coordinate along field lines that plays the role of time. This reparametrization may require that  the particles  be first divided into different classes  depending on their orbit  topology  as is the case,  for a toroidal plasma, of  passing and trapped particles} {(see Ref. \onlinecite{hinton1976theory}).} {
 Finally we stress that a  wide range of application  of the LCS  approach to different plasma configurations is easy to envision.
 For example,  LCSs can offer  a new approach for  the study of anomalous particle transport in space  or astrophysical plasmas  where this  technique  may   complement investigations performed with different tools, see e.g. Ref. \onlinecite{zimbardo2012anomalous} for  heliospheric plasmas.  Conversely, LCS may be looked for in a kinetic plasma description in  on order to identify transport   structures in particle phase space.  An investigation of this type  was  performed in the case of a beam plasma instability in Ref. \onlinecite{carlevaro2015nonlinear} in terms of a one dimensional Vlasov-Poisson system.}

\section*{Acknowledgments}
GDG and DG thanks Dario Borgogno for fruitful discussions.

\bibliography{bibliolcs}

\begin{thebibliography}{48}%
\makeatletter
\providecommand \@ifxundefined [1]{%
 \@ifx{#1\undefined}
}%
\providecommand \@ifnum [1]{%
 \ifnum #1\expandafter \@firstoftwo
 \else \expandafter \@secondoftwo
 \fi
}%
\providecommand \@ifx [1]{%
 \ifx #1\expandafter \@firstoftwo
 \else \expandafter \@secondoftwo
 \fi
}%
\providecommand \natexlab [1]{#1}%
\providecommand \enquote  [1]{``#1''}%
\providecommand \bibnamefont  [1]{#1}%
\providecommand \bibfnamefont [1]{#1}%
\providecommand \citenamefont [1]{#1}%
\providecommand \href@noop [0]{\@secondoftwo}%
\providecommand \href [0]{\begingroup \@sanitize@url \@href}%
\providecommand \@href[1]{\@@startlink{#1}\@@href}%
\providecommand \@@href[1]{\endgroup#1\@@endlink}%
\providecommand \@sanitize@url [0]{\catcode `\\12\catcode `\$12\catcode
  `\&12\catcode `\#12\catcode `\^12\catcode `\_12\catcode `\%12\relax}%
\providecommand \@@startlink[1]{}%
\providecommand \@@endlink[0]{}%
\providecommand \url  [0]{\begingroup\@sanitize@url \@url }%
\providecommand \@url [1]{\endgroup\@href {#1}{\urlprefix }}%
\providecommand \urlprefix  [0]{URL }%
\providecommand \Eprint [0]{\href }%
\providecommand \doibase [0]{http://dx.doi.org/}%
\providecommand \selectlanguage [0]{\@gobble}%
\providecommand \bibinfo  [0]{\@secondoftwo}%
\providecommand \bibfield  [0]{\@secondoftwo}%
\providecommand \translation [1]{[#1]}%
\providecommand \BibitemOpen [0]{}%
\providecommand \bibitemStop [0]{}%
\providecommand \bibitemNoStop [0]{.\EOS\space}%
\providecommand \EOS [0]{\spacefactor3000\relax}%
\providecommand \BibitemShut  [1]{\csname bibitem#1\endcsname}%
\let\auto@bib@innerbib\@empty
\bibitem [{\citenamefont {Haller}\ and\ \citenamefont
  {Yuan}(2000)}]{haller2000lagrangian}%
  \BibitemOpen
  \bibfield  {author} {\bibinfo {author} {\bibfnamefont {G.}~\bibnamefont
  {Haller}}\ and\ \bibinfo {author} {\bibfnamefont {G.}~\bibnamefont {Yuan}},\
  }\href@noop {} {\bibfield  {journal} {\bibinfo  {journal} {Physica D:
  Nonlinear Phenomena}\ }\textbf {\bibinfo {volume} {147}},\ \bibinfo {pages}
  {352} (\bibinfo {year} {2000})}\BibitemShut {NoStop}%
\bibitem [{\citenamefont {Haller}(2015)}]{haller2015lagrangian}%
  \BibitemOpen
  \bibfield  {author} {\bibinfo {author} {\bibfnamefont {G.}~\bibnamefont
  {Haller}},\ }\href@noop {} {\bibfield  {journal} {\bibinfo  {journal} {Annual
  Review of Fluid Mechanics}\ }\textbf {\bibinfo {volume} {47}},\ \bibinfo
  {pages} {137} (\bibinfo {year} {2015})}\BibitemShut {NoStop}%
\bibitem [{\citenamefont {Haller}(2011)}]{haller2011variational}%
  \BibitemOpen
  \bibfield  {author} {\bibinfo {author} {\bibfnamefont {G.}~\bibnamefont
  {Haller}},\ }\href@noop {} {\bibfield  {journal} {\bibinfo  {journal}
  {Physica D: Nonlinear Phenomena}\ }\textbf {\bibinfo {volume} {240}},\
  \bibinfo {pages} {574} (\bibinfo {year} {2011})}\BibitemShut {NoStop}%
\bibitem [{\citenamefont {Coulliette}\ \emph {et~al.}(2007)\citenamefont
  {Coulliette}, \citenamefont {Lekien}, \citenamefont {Paduan}, \citenamefont
  {Haller},\ and\ \citenamefont {Marsden}}]{coulliette2007optimal}%
  \BibitemOpen
  \bibfield  {author} {\bibinfo {author} {\bibfnamefont {C.}~\bibnamefont
  {Coulliette}}, \bibinfo {author} {\bibfnamefont {F.}~\bibnamefont {Lekien}},
  \bibinfo {author} {\bibfnamefont {J.~D.}\ \bibnamefont {Paduan}}, \bibinfo
  {author} {\bibfnamefont {G.}~\bibnamefont {Haller}}, \ and\ \bibinfo {author}
  {\bibfnamefont {J.~E.}\ \bibnamefont {Marsden}},\ }\href@noop {} {\bibfield
  {journal} {\bibinfo  {journal} {Environmental science \& technology}\
  }\textbf {\bibinfo {volume} {41}},\ \bibinfo {pages} {6562} (\bibinfo {year}
  {2007})}\BibitemShut {NoStop}%
\bibitem [{\citenamefont {Shadden}\ and\ \citenamefont
  {Taylor}(2008)}]{shadden2008characterization}%
  \BibitemOpen
  \bibfield  {author} {\bibinfo {author} {\bibfnamefont {S.~C.}\ \bibnamefont
  {Shadden}}\ and\ \bibinfo {author} {\bibfnamefont {C.~A.}\ \bibnamefont
  {Taylor}},\ }\href@noop {} {\bibfield  {journal} {\bibinfo  {journal} {Annals
  of biomedical engineering}\ }\textbf {\bibinfo {volume} {36}},\ \bibinfo
  {pages} {1152} (\bibinfo {year} {2008})}\BibitemShut {NoStop}%
\bibitem [{\citenamefont {Huhn}\ \emph {et~al.}(2012)\citenamefont {Huhn},
  \citenamefont {Kameke}, \citenamefont {P{\'e}rez-Mu{\~n}uzuri}, \citenamefont
  {Olascoaga},\ and\ \citenamefont {Beron-Vera}}]{huhn2012impact}%
  \BibitemOpen
  \bibfield  {author} {\bibinfo {author} {\bibfnamefont {F.}~\bibnamefont
  {Huhn}}, \bibinfo {author} {\bibfnamefont {A.}~\bibnamefont {Kameke}},
  \bibinfo {author} {\bibfnamefont {V.}~\bibnamefont {P{\'e}rez-Mu{\~n}uzuri}},
  \bibinfo {author} {\bibfnamefont {M.}~\bibnamefont {Olascoaga}}, \ and\
  \bibinfo {author} {\bibfnamefont {F.}~\bibnamefont {Beron-Vera}},\
  }\href@noop {} {\bibfield  {journal} {\bibinfo  {journal} {Geophysical
  Research Letters}\ }\textbf {\bibinfo {volume} {39}} (\bibinfo {year}
  {2012})}\BibitemShut {NoStop}%
\bibitem [{\citenamefont {Peng}\ and\ \citenamefont
  {Dabiri}(2009)}]{peng2009transport}%
  \BibitemOpen
  \bibfield  {author} {\bibinfo {author} {\bibfnamefont {J.}~\bibnamefont
  {Peng}}\ and\ \bibinfo {author} {\bibfnamefont {J.}~\bibnamefont {Dabiri}},\
  }\href@noop {} {\bibfield  {journal} {\bibinfo  {journal} {Journal of Fluid
  Mechanics}\ }\textbf {\bibinfo {volume} {623}},\ \bibinfo {pages} {75}
  (\bibinfo {year} {2009})}\BibitemShut {NoStop}%
\bibitem [{\citenamefont {Tang}\ \emph {et~al.}(2010)\citenamefont {Tang},
  \citenamefont {Mathur}, \citenamefont {Haller}, \citenamefont {Hahn},\ and\
  \citenamefont {Ruggiero}}]{tang2010lagrangian}%
  \BibitemOpen
  \bibfield  {author} {\bibinfo {author} {\bibfnamefont {W.}~\bibnamefont
  {Tang}}, \bibinfo {author} {\bibfnamefont {M.}~\bibnamefont {Mathur}},
  \bibinfo {author} {\bibfnamefont {G.}~\bibnamefont {Haller}}, \bibinfo
  {author} {\bibfnamefont {D.~C.}\ \bibnamefont {Hahn}}, \ and\ \bibinfo
  {author} {\bibfnamefont {F.~H.}\ \bibnamefont {Ruggiero}},\ }\href@noop {}
  {\bibfield  {journal} {\bibinfo  {journal} {Journal of the Atmospheric
  Sciences}\ }\textbf {\bibinfo {volume} {67}},\ \bibinfo {pages} {2307}
  (\bibinfo {year} {2010})}\BibitemShut {NoStop}%
\bibitem [{\citenamefont {Carlevaro}\ \emph {et~al.}(2015)\citenamefont
  {Carlevaro}, \citenamefont {Falessi}, \citenamefont {Montani},\ and\
  \citenamefont {Zonca}}]{carlevaro2015nonlinear}%
  \BibitemOpen
  \bibfield  {author} {\bibinfo {author} {\bibfnamefont {N.}~\bibnamefont
  {Carlevaro}}, \bibinfo {author} {\bibfnamefont {M.~V.}\ \bibnamefont
  {Falessi}}, \bibinfo {author} {\bibfnamefont {G.}~\bibnamefont {Montani}}, \
  and\ \bibinfo {author} {\bibfnamefont {F.}~\bibnamefont {Zonca}},\
  }\href@noop {} {\bibfield  {journal} {\bibinfo  {journal} {Journal of Plasma
  Physics}\ }\textbf {\bibinfo {volume} {81}},\ \bibinfo {pages} {495810515}
  (\bibinfo {year} {2015})}\BibitemShut {NoStop}%
\bibitem [{\citenamefont {Chian}\ \emph {et~al.}(2014)\citenamefont {Chian},
  \citenamefont {Rempel}, \citenamefont {Aulanier}, \citenamefont {Schmieder},
  \citenamefont {Shadden}, \citenamefont {Welsch},\ and\ \citenamefont
  {Yeates}}]{chian2014detection}%
  \BibitemOpen
  \bibfield  {author} {\bibinfo {author} {\bibfnamefont {A.~C.-L.}\
  \bibnamefont {Chian}}, \bibinfo {author} {\bibfnamefont {E.~L.}\ \bibnamefont
  {Rempel}}, \bibinfo {author} {\bibfnamefont {G.}~\bibnamefont {Aulanier}},
  \bibinfo {author} {\bibfnamefont {B.}~\bibnamefont {Schmieder}}, \bibinfo
  {author} {\bibfnamefont {S.~C.}\ \bibnamefont {Shadden}}, \bibinfo {author}
  {\bibfnamefont {B.~T.}\ \bibnamefont {Welsch}}, \ and\ \bibinfo {author}
  {\bibfnamefont {A.~R.}\ \bibnamefont {Yeates}},\ }\href@noop {} {\bibfield
  {journal} {\bibinfo  {journal} {The Astrophysical Journal}\ }\textbf
  {\bibinfo {volume} {786}},\ \bibinfo {pages} {51} (\bibinfo {year}
  {2014})}\BibitemShut {NoStop}%
\bibitem [{\citenamefont {Rempel}\ \emph {et~al.}(2013)\citenamefont {Rempel},
  \citenamefont {Chian}, \citenamefont {Brandenburg}, \citenamefont
  {Mu{\~n}oz},\ and\ \citenamefont {Shadden}}]{rempel2013coherent}%
  \BibitemOpen
  \bibfield  {author} {\bibinfo {author} {\bibfnamefont {E.~L.}\ \bibnamefont
  {Rempel}}, \bibinfo {author} {\bibfnamefont {A.~C.-L.}\ \bibnamefont
  {Chian}}, \bibinfo {author} {\bibfnamefont {A.}~\bibnamefont {Brandenburg}},
  \bibinfo {author} {\bibfnamefont {P.~R.}\ \bibnamefont {Mu{\~n}oz}}, \ and\
  \bibinfo {author} {\bibfnamefont {S.~C.}\ \bibnamefont {Shadden}},\
  }\href@noop {} {\bibfield  {journal} {\bibinfo  {journal} {Journal of Fluid
  Mechanics}\ }\textbf {\bibinfo {volume} {729}},\ \bibinfo {pages} {309}
  (\bibinfo {year} {2013})}\BibitemShut {NoStop}%
\bibitem [{\citenamefont {Shadden}, \citenamefont {Lekien},\ and\ \citenamefont
  {Marsden}(2005)}]{shadden2005definition}%
  \BibitemOpen
  \bibfield  {author} {\bibinfo {author} {\bibfnamefont {S.~C.}\ \bibnamefont
  {Shadden}}, \bibinfo {author} {\bibfnamefont {F.}~\bibnamefont {Lekien}}, \
  and\ \bibinfo {author} {\bibfnamefont {J.~E.}\ \bibnamefont {Marsden}},\
  }\href@noop {} {\bibfield  {journal} {\bibinfo  {journal} {Physica D:
  Nonlinear Phenomena}\ }\textbf {\bibinfo {volume} {212}},\ \bibinfo {pages}
  {271} (\bibinfo {year} {2005})}\BibitemShut {NoStop}%
\bibitem [{\citenamefont {Farazmand}\ and\ \citenamefont
  {Haller}(2012)}]{farazmand2012computing}%
  \BibitemOpen
  \bibfield  {author} {\bibinfo {author} {\bibfnamefont {M.}~\bibnamefont
  {Farazmand}}\ and\ \bibinfo {author} {\bibfnamefont {G.}~\bibnamefont
  {Haller}},\ }\href@noop {} {\bibfield  {journal} {\bibinfo  {journal} {Chaos:
  An Interdisciplinary Journal of Nonlinear Science}\ }\textbf {\bibinfo
  {volume} {22}},\ \bibinfo {pages} {013128} (\bibinfo {year}
  {2012})}\BibitemShut {NoStop}%
\bibitem [{\citenamefont {Onu}, \citenamefont {Huhn},\ and\ \citenamefont
  {Haller}(2015)}]{onu2015lcs}%
  \BibitemOpen
  \bibfield  {author} {\bibinfo {author} {\bibfnamefont {K.}~\bibnamefont
  {Onu}}, \bibinfo {author} {\bibfnamefont {F.}~\bibnamefont {Huhn}}, \ and\
  \bibinfo {author} {\bibfnamefont {G.}~\bibnamefont {Haller}},\ }\href@noop {}
  {\bibfield  {journal} {\bibinfo  {journal} {Journal of Computational
  Science}\ }\textbf {\bibinfo {volume} {7}},\ \bibinfo {pages} {26} (\bibinfo
  {year} {2015})}\BibitemShut {NoStop}%
\bibitem [{\citenamefont {Borgogno}\ \emph {et~al.}(2011)\citenamefont
  {Borgogno}, \citenamefont {Grasso}, \citenamefont {Pegoraro},\ and\
  \citenamefont {Schep}}]{borgogno2011barriers}%
  \BibitemOpen
  \bibfield  {author} {\bibinfo {author} {\bibfnamefont {D.}~\bibnamefont
  {Borgogno}}, \bibinfo {author} {\bibfnamefont {D.}~\bibnamefont {Grasso}},
  \bibinfo {author} {\bibfnamefont {F.}~\bibnamefont {Pegoraro}}, \ and\
  \bibinfo {author} {\bibfnamefont {T.}~\bibnamefont {Schep}},\ }\href@noop {}
  {\bibfield  {journal} {\bibinfo  {journal} {Physics of Plasmas
  (1994-present)}\ }\textbf {\bibinfo {volume} {18}},\ \bibinfo {pages}
  {102307} (\bibinfo {year} {2011})}\BibitemShut {NoStop}%
\bibitem [{\citenamefont {Rubino}\ \emph {et~al.}(2015)\citenamefont {Rubino},
  \citenamefont {Borgogno}, \citenamefont {Veranda}, \citenamefont {Bonfiglio},
  \citenamefont {Cappello},\ and\ \citenamefont
  {Grasso}}]{rubino2015detection}%
  \BibitemOpen
  \bibfield  {author} {\bibinfo {author} {\bibfnamefont {G.}~\bibnamefont
  {Rubino}}, \bibinfo {author} {\bibfnamefont {D.}~\bibnamefont {Borgogno}},
  \bibinfo {author} {\bibfnamefont {M.}~\bibnamefont {Veranda}}, \bibinfo
  {author} {\bibfnamefont {D.}~\bibnamefont {Bonfiglio}}, \bibinfo {author}
  {\bibfnamefont {S.}~\bibnamefont {Cappello}}, \ and\ \bibinfo {author}
  {\bibfnamefont {D.}~\bibnamefont {Grasso}},\ }\href@noop {} {\bibfield
  {journal} {\bibinfo  {journal} {Plasma Physics and Controlled Fusion}\
  }\textbf {\bibinfo {volume} {57}},\ \bibinfo {pages} {085004} (\bibinfo
  {year} {2015})}\BibitemShut {NoStop}%
\bibitem [{\citenamefont {Falessi}, \citenamefont {Pegoraro},\ and\
  \citenamefont {Schep}(2015)}]{falessi2015lagrangian}%
  \BibitemOpen
  \bibfield  {author} {\bibinfo {author} {\bibfnamefont {M.}~\bibnamefont
  {Falessi}}, \bibinfo {author} {\bibfnamefont {F.}~\bibnamefont {Pegoraro}}, \
  and\ \bibinfo {author} {\bibfnamefont {T.}~\bibnamefont {Schep}},\
  }\href@noop {} {\bibfield  {journal} {\bibinfo  {journal} {Journal of Plasma
  Physics}\ }\textbf {\bibinfo {volume} {81}},\ \bibinfo {pages} {495810505}
  (\bibinfo {year} {2015})}\BibitemShut {NoStop}%
\bibitem [{\citenamefont {Borgogno}, \citenamefont {Perona},\ and\
  \citenamefont {Grasso}(2017)}]{borgogno17}%
  \BibitemOpen
  \bibfield  {author} {\bibinfo {author} {\bibfnamefont {D.}~\bibnamefont
  {Borgogno}}, \bibinfo {author} {\bibfnamefont {A.}~\bibnamefont {Perona}}, \
  and\ \bibinfo {author} {\bibfnamefont {D.}~\bibnamefont {Grasso}},\
  }\href@noop {} {\bibfield  {journal} {\bibinfo  {journal} {Physics of Plasmas
  (1994-present)}\ }\textbf {\bibinfo {volume} {accepted for publication on}}
  (\bibinfo {year} {2017})}\BibitemShut {NoStop}%
\bibitem [{\citenamefont {Borgogno}\ \emph {et~al.}(2008)\citenamefont
  {Borgogno}, \citenamefont {Grasso}, \citenamefont {Pegoraro},\ and\
  \citenamefont {Schep}}]{borgogno2008stable}%
  \BibitemOpen
  \bibfield  {author} {\bibinfo {author} {\bibfnamefont {D.}~\bibnamefont
  {Borgogno}}, \bibinfo {author} {\bibfnamefont {D.}~\bibnamefont {Grasso}},
  \bibinfo {author} {\bibfnamefont {F.}~\bibnamefont {Pegoraro}}, \ and\
  \bibinfo {author} {\bibfnamefont {T.}~\bibnamefont {Schep}},\ }\href@noop {}
  {\bibfield  {journal} {\bibinfo  {journal} {Physics of Plasmas
  (1994-present)}\ }\textbf {\bibinfo {volume} {15}},\ \bibinfo {pages}
  {102308} (\bibinfo {year} {2008})}\BibitemShut {NoStop}%
\bibitem [{\citenamefont {Cary}\ and\ \citenamefont
  {Littlejohn}(1983)}]{cary1983noncanonical}%
  \BibitemOpen
  \bibfield  {author} {\bibinfo {author} {\bibfnamefont {J.~R.}\ \bibnamefont
  {Cary}}\ and\ \bibinfo {author} {\bibfnamefont {R.~G.}\ \bibnamefont
  {Littlejohn}},\ }\href@noop {} {\bibfield  {journal} {\bibinfo  {journal}
  {Annals of Physics}\ }\textbf {\bibinfo {volume} {151}},\ \bibinfo {pages}
  {1} (\bibinfo {year} {1983})}\BibitemShut {NoStop}%
\bibitem [{\citenamefont {Kruskal}(1952)}]{kruskal1952some}%
  \BibitemOpen
  \bibfield  {author} {\bibinfo {author} {\bibfnamefont {M.~D.}\ \bibnamefont
  {Kruskal}},\ }\href@noop {} {\enquote {\bibinfo {title} {Some properties of
  rotational transforms},}\ }\bibinfo {type} {Tech. Rep.}\ (\bibinfo
  {institution} {Forrestal Research Center, Princeton Univ.},\ \bibinfo {year}
  {1952})\BibitemShut {NoStop}%
\bibitem [{\citenamefont {Kerst}(1962)}]{kerst1962influence}%
  \BibitemOpen
  \bibfield  {author} {\bibinfo {author} {\bibfnamefont {D.}~\bibnamefont
  {Kerst}},\ }\href@noop {} {\bibfield  {journal} {\bibinfo  {journal} {Journal
  of Nuclear Energy. Part C, Plasma Physics, Accelerators, Thermonuclear
  Research}\ }\textbf {\bibinfo {volume} {4}},\ \bibinfo {pages} {253}
  (\bibinfo {year} {1962})}\BibitemShut {NoStop}%
\bibitem [{\citenamefont {Gelfand}\ \emph {et~al.}(1962)\citenamefont
  {Gelfand}, \citenamefont {Morozov}, \citenamefont {Sololev}, \citenamefont
  {Graev},\ and\ \citenamefont {Zueva}}]{gelfand1962magnetic}%
  \BibitemOpen
  \bibfield  {author} {\bibinfo {author} {\bibfnamefont {I.}~\bibnamefont
  {Gelfand}}, \bibinfo {author} {\bibfnamefont {A.}~\bibnamefont {Morozov}},
  \bibinfo {author} {\bibfnamefont {L.}~\bibnamefont {Sololev}}, \bibinfo
  {author} {\bibfnamefont {M.}~\bibnamefont {Graev}}, \ and\ \bibinfo {author}
  {\bibfnamefont {N.}~\bibnamefont {Zueva}},\ }\href@noop {} {\bibfield
  {journal} {\bibinfo  {journal} {Soviet physics-technical physics}\ }\textbf
  {\bibinfo {volume} {6}},\ \bibinfo {pages} {852} (\bibinfo {year}
  {1962})}\BibitemShut {NoStop}%
\bibitem [{\citenamefont {Morozov}\ and\ \citenamefont
  {Solov'ev}(1966)}]{morozov1966structure}%
  \BibitemOpen
  \bibfield  {author} {\bibinfo {author} {\bibfnamefont {A.}~\bibnamefont
  {Morozov}}\ and\ \bibinfo {author} {\bibfnamefont {L.}~\bibnamefont
  {Solov'ev}},\ }\href@noop {} {\bibfield  {journal} {\bibinfo  {journal}
  {Reviews of Plasma Physics}\ }\textbf {\bibinfo {volume} {2}},\ \bibinfo
  {pages} {1} (\bibinfo {year} {1966})}\BibitemShut {NoStop}%
\bibitem [{\citenamefont {Boozer}(1981)}]{boozer1981plasma}%
  \BibitemOpen
  \bibfield  {author} {\bibinfo {author} {\bibfnamefont {A.~H.}\ \bibnamefont
  {Boozer}},\ }\href@noop {} {\bibfield  {journal} {\bibinfo  {journal}
  {Physics of Fluids (1958-1988)}\ }\textbf {\bibinfo {volume} {24}},\ \bibinfo
  {pages} {1999} (\bibinfo {year} {1981})}\BibitemShut {NoStop}%
\bibitem [{\citenamefont {Rosenbluth}\ \emph {et~al.}(1966)\citenamefont
  {Rosenbluth}, \citenamefont {Sagdeev}, \citenamefont {Taylor},\ and\
  \citenamefont {Zaslavski}}]{rosen1}%
  \BibitemOpen
  \bibfield  {author} {\bibinfo {author} {\bibfnamefont {M.}~\bibnamefont
  {Rosenbluth}}, \bibinfo {author} {\bibfnamefont {R.}~\bibnamefont {Sagdeev}},
  \bibinfo {author} {\bibfnamefont {J.}~\bibnamefont {Taylor}}, \ and\ \bibinfo
  {author} {\bibfnamefont {G.}~\bibnamefont {Zaslavski}},\ }\href@noop {}
  {\bibfield  {journal} {\bibinfo  {journal} {Nuclear Fusion}\ }\textbf
  {\bibinfo {volume} {6}},\ \bibinfo {pages} {297} (\bibinfo {year}
  {1966})}\BibitemShut {NoStop}%
\bibitem [{\citenamefont {Elsasser}(1986)}]{elsasser}%
  \BibitemOpen
  \bibfield  {author} {\bibinfo {author} {\bibfnamefont {K.}~\bibnamefont
  {Elsasser}},\ }\href@noop {} {\bibfield  {journal} {\bibinfo  {journal}
  {Plasma physics and controlled fusion}\ }\textbf {\bibinfo {volume} {28}},\
  \bibinfo {pages} {1743} (\bibinfo {year} {1986})}\BibitemShut {NoStop}%
\bibitem [{\citenamefont {Morrison}(2000)}]{morrison}%
  \BibitemOpen
  \bibfield  {author} {\bibinfo {author} {\bibfnamefont {P.}~\bibnamefont
  {Morrison}},\ }\href@noop {} {\bibfield  {journal} {\bibinfo  {journal}
  {Physics of Plasmas}\ }\textbf {\bibinfo {volume} {7}},\ \bibinfo {pages}
  {2279} (\bibinfo {year} {2000})}\BibitemShut {NoStop}%
\bibitem [{\citenamefont {Boozer}(2005)}]{boozer2005physics}%
  \BibitemOpen
  \bibfield  {author} {\bibinfo {author} {\bibfnamefont {A.~H.}\ \bibnamefont
  {Boozer}},\ }\href@noop {} {\bibfield  {journal} {\bibinfo  {journal}
  {Reviews of modern physics}\ }\textbf {\bibinfo {volume} {76}},\ \bibinfo
  {pages} {1071} (\bibinfo {year} {2005})}\BibitemShut {NoStop}%
\bibitem [{\citenamefont {Rechester}\ and\ \citenamefont
  {Rosenbluth}(1978)}]{rosen2}%
  \BibitemOpen
  \bibfield  {author} {\bibinfo {author} {\bibfnamefont {A.}~\bibnamefont
  {Rechester}}\ and\ \bibinfo {author} {\bibfnamefont {M.}~\bibnamefont
  {Rosenbluth}},\ }\href@noop {} {\bibfield  {journal} {\bibinfo  {journal}
  {Physical Review Letters}\ }\textbf {\bibinfo {volume} {40}},\ \bibinfo
  {pages} {38} (\bibinfo {year} {1978})}\BibitemShut {NoStop}%
\bibitem [{\citenamefont {Arnold}, \citenamefont {Kozlov},\ and\ \citenamefont
  {Neishtadt}(2007)}]{arnold2007mathematical}%
  \BibitemOpen
  \bibfield  {author} {\bibinfo {author} {\bibfnamefont {V.~I.}\ \bibnamefont
  {Arnold}}, \bibinfo {author} {\bibfnamefont {V.~V.}\ \bibnamefont {Kozlov}},
  \ and\ \bibinfo {author} {\bibfnamefont {A.~I.}\ \bibnamefont {Neishtadt}},\
  }\href@noop {} {\emph {\bibinfo {title} {Mathematical aspects of classical
  and celestial mechanics}}},\ Vol.~\bibinfo {volume} {3}\ (\bibinfo
  {publisher} {Springer Science \& Business Media},\ \bibinfo {year}
  {2007})\BibitemShut {NoStop}%
\bibitem [{\citenamefont {Wiggins}(1992{\natexlab{a}})}]{wiggins1992chaotic}%
  \BibitemOpen
  \bibfield  {author} {\bibinfo {author} {\bibfnamefont {S.}~\bibnamefont
  {Wiggins}},\ }\href@noop {} {\bibfield  {journal} {\bibinfo  {journal} {NASA
  STI/Recon Technical Report A}\ }\textbf {\bibinfo {volume} {92}},\ \bibinfo
  {pages} {28228} (\bibinfo {year} {1992}{\natexlab{a}})}\BibitemShut {NoStop}%
\bibitem [{\citenamefont {Cencini}, \citenamefont {Cecconi},\ and\
  \citenamefont {Vulpiani}(2010)}]{cencini2010simple}%
  \BibitemOpen
  \bibfield  {author} {\bibinfo {author} {\bibfnamefont {M.}~\bibnamefont
  {Cencini}}, \bibinfo {author} {\bibfnamefont {F.}~\bibnamefont {Cecconi}}, \
  and\ \bibinfo {author} {\bibfnamefont {A.}~\bibnamefont {Vulpiani}},\
  }\href@noop {} {\bibfield  {journal} {\bibinfo  {journal} {Chaos}\ }\textbf
  {\bibinfo {volume} {10}},\ \bibinfo {pages} {9789814277662\_0001} (\bibinfo
  {year} {2010})}\BibitemShut {NoStop}%
\bibitem [{\citenamefont {Mackay}, \citenamefont {Meiss},\ and\ \citenamefont
  {Percival}(1987)}]{mackay1987resonances}%
  \BibitemOpen
  \bibfield  {author} {\bibinfo {author} {\bibfnamefont {R.~S.}\ \bibnamefont
  {Mackay}}, \bibinfo {author} {\bibfnamefont {J.}~\bibnamefont {Meiss}}, \
  and\ \bibinfo {author} {\bibfnamefont {I.}~\bibnamefont {Percival}},\
  }\href@noop {} {\bibfield  {journal} {\bibinfo  {journal} {Physica D:
  Nonlinear Phenomena}\ }\textbf {\bibinfo {volume} {27}},\ \bibinfo {pages}
  {1} (\bibinfo {year} {1987})}\BibitemShut {NoStop}%
\bibitem [{\citenamefont {Meiss}(2015)}]{meiss2015thirty}%
  \BibitemOpen
  \bibfield  {author} {\bibinfo {author} {\bibfnamefont {J.}~\bibnamefont
  {Meiss}},\ }\href@noop {} {\bibfield  {journal} {\bibinfo  {journal} {Chaos:
  An Interdisciplinary Journal of Nonlinear Science}\ }\textbf {\bibinfo
  {volume} {25}},\ \bibinfo {pages} {097602} (\bibinfo {year}
  {2015})}\BibitemShut {NoStop}%
\bibitem [{\citenamefont {Ottino}(1989)}]{ottino1989kinematics}%
  \BibitemOpen
  \bibfield  {author} {\bibinfo {author} {\bibfnamefont {J.~M.}\ \bibnamefont
  {Ottino}},\ }\href@noop {} {\emph {\bibinfo {title} {The kinematics of
  mixing: stretching, chaos, and transport}}},\ Vol.~\bibinfo {volume} {3}\
  (\bibinfo  {publisher} {Cambridge university press},\ \bibinfo {year}
  {1989})\BibitemShut {NoStop}%
\bibitem [{\citenamefont {Wiggins}(2013)}]{wiggins2013chaotic}%
  \BibitemOpen
  \bibfield  {author} {\bibinfo {author} {\bibfnamefont {S.}~\bibnamefont
  {Wiggins}},\ }\href@noop {} {\emph {\bibinfo {title} {Chaotic transport in
  dynamical systems}}},\ Vol.~\bibinfo {volume} {2}\ (\bibinfo  {publisher}
  {Springer Science \& Business Media},\ \bibinfo {year} {2013})\BibitemShut
  {NoStop}%
\bibitem [{\citenamefont {Rom-Kedar}\ and\ \citenamefont
  {Wiggins}(1990)}]{rom1990transport}%
  \BibitemOpen
  \bibfield  {author} {\bibinfo {author} {\bibfnamefont {V.}~\bibnamefont
  {Rom-Kedar}}\ and\ \bibinfo {author} {\bibfnamefont {S.}~\bibnamefont
  {Wiggins}},\ }\href@noop {} {\bibfield  {journal} {\bibinfo  {journal}
  {Archive for Rational Mechanics and Analysis}\ }\textbf {\bibinfo {volume}
  {109}},\ \bibinfo {pages} {239} (\bibinfo {year} {1990})}\BibitemShut
  {NoStop}%
\bibitem [{\citenamefont {Rom-Kedar}, \citenamefont {Leonard},\ and\
  \citenamefont {Wiggins}(1990)}]{rom1990analytical}%
  \BibitemOpen
  \bibfield  {author} {\bibinfo {author} {\bibfnamefont {V.}~\bibnamefont
  {Rom-Kedar}}, \bibinfo {author} {\bibfnamefont {A.}~\bibnamefont {Leonard}},
  \ and\ \bibinfo {author} {\bibfnamefont {S.}~\bibnamefont {Wiggins}},\
  }\href@noop {} {\bibfield  {journal} {\bibinfo  {journal} {Journal of Fluid
  Mechanics}\ }\textbf {\bibinfo {volume} {214}},\ \bibinfo {pages} {347}
  (\bibinfo {year} {1990})}\BibitemShut {NoStop}%
\bibitem [{\citenamefont {Malhotra}\ and\ \citenamefont
  {Wiggins}(1998)}]{malhotra1998geometric}%
  \BibitemOpen
  \bibfield  {author} {\bibinfo {author} {\bibfnamefont {N.}~\bibnamefont
  {Malhotra}}\ and\ \bibinfo {author} {\bibfnamefont {S.}~\bibnamefont
  {Wiggins}},\ }\href@noop {} {\bibfield  {journal} {\bibinfo  {journal}
  {Journal of nonlinear science}\ }\textbf {\bibinfo {volume} {8}},\ \bibinfo
  {pages} {401} (\bibinfo {year} {1998})}\BibitemShut {NoStop}%
\bibitem [{\citenamefont {Wiggins}(1992{\natexlab{b}})}]{wwiggins1992chaotic}%
  \BibitemOpen
  \bibfield  {author} {\bibinfo {author} {\bibfnamefont {S.}~\bibnamefont
  {Wiggins}},\ }\href@noop {} {\bibfield  {journal} {\bibinfo  {journal} {NASA
  STI/Recon Technical Report A}\ }\textbf {\bibinfo {volume} {92}},\ \bibinfo
  {pages} {28228} (\bibinfo {year} {1992}{\natexlab{b}})}\BibitemShut {NoStop}%
\bibitem [{\citenamefont {Chen}(1987)}]{chen1987area}%
  \BibitemOpen
  \bibfield  {author} {\bibinfo {author} {\bibfnamefont {Q.}~\bibnamefont
  {Chen}},\ }\href@noop {} {\bibfield  {journal} {\bibinfo  {journal} {Physics
  Letters A}\ }\textbf {\bibinfo {volume} {123}},\ \bibinfo {pages} {444}
  (\bibinfo {year} {1987})}\BibitemShut {NoStop}%
\bibitem [{\citenamefont {Raynal}\ and\ \citenamefont
  {Wiggins}(2006)}]{raynal2006lobe}%
  \BibitemOpen
  \bibfield  {author} {\bibinfo {author} {\bibfnamefont {F.}~\bibnamefont
  {Raynal}}\ and\ \bibinfo {author} {\bibfnamefont {S.}~\bibnamefont
  {Wiggins}},\ }\href@noop {} {\bibfield  {journal} {\bibinfo  {journal}
  {Physica D: Nonlinear Phenomena}\ }\textbf {\bibinfo {volume} {223}},\
  \bibinfo {pages} {7} (\bibinfo {year} {2006})}\BibitemShut {NoStop}%
\bibitem [{\citenamefont {Borgogno}\ \emph {et~al.}(2005)\citenamefont
  {Borgogno}, \citenamefont {Grasso}, \citenamefont {Porcelli}, \citenamefont
  {Califano}, \citenamefont {Pegoraro},\ and\ \citenamefont
  {Farina}}]{borgogno2005aspects}%
  \BibitemOpen
  \bibfield  {author} {\bibinfo {author} {\bibfnamefont {D.}~\bibnamefont
  {Borgogno}}, \bibinfo {author} {\bibfnamefont {D.}~\bibnamefont {Grasso}},
  \bibinfo {author} {\bibfnamefont {F.}~\bibnamefont {Porcelli}}, \bibinfo
  {author} {\bibfnamefont {F.}~\bibnamefont {Califano}}, \bibinfo {author}
  {\bibfnamefont {F.}~\bibnamefont {Pegoraro}}, \ and\ \bibinfo {author}
  {\bibfnamefont {D.}~\bibnamefont {Farina}},\ }\href@noop {} {\bibfield
  {journal} {\bibinfo  {journal} {Physics of plasmas}\ }\textbf {\bibinfo
  {volume} {12}},\ \bibinfo {pages} {032309} (\bibinfo {year}
  {2005})}\BibitemShut {NoStop}%
\bibitem [{\citenamefont {Frieman}\ and\ \citenamefont
  {Chen}(1982)}]{frieman1982nonlinear}%
  \BibitemOpen
  \bibfield  {author} {\bibinfo {author} {\bibfnamefont {E.}~\bibnamefont
  {Frieman}}\ and\ \bibinfo {author} {\bibfnamefont {L.}~\bibnamefont {Chen}},\
  }\href@noop {} {\bibfield  {journal} {\bibinfo  {journal} {Physics of Fluids
  (1958-1988)}\ }\textbf {\bibinfo {volume} {25}},\ \bibinfo {pages} {502}
  (\bibinfo {year} {1982})}\BibitemShut {NoStop}%
\bibitem [{\citenamefont {Brizard}\ and\ \citenamefont
  {Hahm}(2007)}]{brizard2007foundations}%
  \BibitemOpen
  \bibfield  {author} {\bibinfo {author} {\bibfnamefont {A.}~\bibnamefont
  {Brizard}}\ and\ \bibinfo {author} {\bibfnamefont {T.}~\bibnamefont {Hahm}},\
  }\href@noop {} {\bibfield  {journal} {\bibinfo  {journal} {Reviews of modern
  physics}\ }\textbf {\bibinfo {volume} {79}},\ \bibinfo {pages} {421}
  (\bibinfo {year} {2007})}\BibitemShut {NoStop}%
\bibitem [{\citenamefont {Hinton}\ and\ \citenamefont
  {Hazeltine}(1976)}]{hinton1976theory}%
  \BibitemOpen
  \bibfield  {author} {\bibinfo {author} {\bibfnamefont {F.}~\bibnamefont
  {Hinton}}\ and\ \bibinfo {author} {\bibfnamefont {R.}~\bibnamefont
  {Hazeltine}},\ }\href@noop {} {\bibfield  {journal} {\bibinfo  {journal}
  {Reviews of Modern Physics}\ }\textbf {\bibinfo {volume} {48}},\ \bibinfo
  {pages} {239} (\bibinfo {year} {1976})}\BibitemShut {NoStop}%
\bibitem [{\citenamefont {Zimbardo}\ \emph {et~al.}(2012)\citenamefont
  {Zimbardo}, \citenamefont {Perri}, \citenamefont {Pommois},\ and\
  \citenamefont {Veltri}}]{zimbardo2012anomalous}%
  \BibitemOpen
  \bibfield  {author} {\bibinfo {author} {\bibfnamefont {G.}~\bibnamefont
  {Zimbardo}}, \bibinfo {author} {\bibfnamefont {S.}~\bibnamefont {Perri}},
  \bibinfo {author} {\bibfnamefont {P.}~\bibnamefont {Pommois}}, \ and\
  \bibinfo {author} {\bibfnamefont {P.}~\bibnamefont {Veltri}},\ }\href@noop {}
  {\bibfield  {journal} {\bibinfo  {journal} {Advances in Space Research}\
  }\textbf {\bibinfo {volume} {49}},\ \bibinfo {pages} {1633} (\bibinfo {year}
  {2012})}\BibitemShut {NoStop}%
\end{thebibliography}%

\end{document}